\documentclass[10pt,conference]{IEEEtran}
\IEEEoverridecommandlockouts
\usepackage{amsmath,amssymb,amsfonts}
\usepackage{algorithmic}
\usepackage{graphicx}
\usepackage{textcomp}
\usepackage{xcolor}
\usepackage{subfigure}
\usepackage{hyperref}

\usepackage[style=ieee, maxnames=1, minnames=1]{biblatex}
\newif\ifcolormarker
\colormarkerfalse

\newcommand{\rrm}[1]{\ifcolormarker
\textcolor{red}{#1}%
\else
#1%
\fi } 

\usepackage{xspace}
\newcommand{\eg}{\textit{e.g.},\xspace}
\newcommand{\ie}{\textit{i.e.},\xspace}
\newcommand{\name}{\textsc{GraphWave}\xspace}

\begin{document}

\title{From Waves to Graphs: A Ray-Tracing-Inspired Neural Radio Propagation Model\\
}

\author{\IEEEauthorblockN{Paul~Almasan\IEEEauthorrefmark{2},
Stefanos~Bakirtzis\IEEEauthorrefmark{3}\thanks{This work was done while Stefanos Bakirtzis was with Telefónica Research.}, José~Suárez-Varela\IEEEauthorrefmark{2}, Andra~Lutu\IEEEauthorrefmark{2}\\
\IEEEauthorrefmark{2}Telefónica Research, \IEEEauthorrefmark{3}University of Cambridge\\
}}
\maketitle

\begin{abstract}

Artificial intelligence-driven radio propagation models provide agile and robust solutions for mobile network operators  in their effort to ensure the optimal performance of the wireless ecosystem and support its efficient expansion. In this paper, we introduce \name, a neural \textit{graph-driven} propagation solver hinging on the governing principles of ray tracing. The proposed model leverages a digitized version of the propagation environment to build a point cloud and extract an equivalent graph representation of the radio environment. By applying neural message passing over the equivalent graph, it allows the model to accurately infer radio-related quantities, \eg received signal strength, in a three-dimensional environment. We showcase the use of \name as a radio environment digital twin and we demonstrate that the model can learn from synthetic and real-world data while achieving low inference times. 

\end{abstract}


\section{Introduction}

Artificial intelligence (AI) has transformed various aspects of the wireless ecosystem along with the methods and tools used to underpin its effective operation \cite{6G_AI}. In particular, AI has enabled the emergence of radio access network (RAN) digital twins (DT) that provide predictive and data-driven representations of the propagation conditions experienced by the network. Notably, a drastic shift has been observed in radio propagation modeling, moving from conventional empirical or deterministic models to AI-powered solutions that exhibit enhanced accuracy,  computational efficiency, and lower cost~\cite{AI_Prop}. Such models yield an invaluable tool for mobile network operators (MNOs) to support predictive control and optimization of the RAN, enabling critical applications such as network planning, coverage optimization, interference management, beamforming, and localization \cite{Empowering_DL_Prop_Models}. 

To mold AI-powered radio propagation tools, one needs a model that accurately translates data from a radio propagation scene---with some transmitting devices deployed---into some radio quantities of interest (QoI), typically the received signal strength (RSS) or the path loss (PL). To this end, it is crucial to properly co-design a data representation that captures the fundamental factors influencing radio propagation in the scene, along with an AI model that can effectively learn and abstract patterns from these representations. Early works in this direction use tabular data along with multi-layer perceptrons (MLPs)~\cite{Outdoor_ANN, Indoor_ANN}. Later works use two-dimensional (2D) image-like visualizations of the propagation environment and other physics-based input features, which can be combined with convolutional neural networks \cite{CNN_Propagation, Radio_UNet, EM_DeepRay} or transformer-based encoding-decoding architectures \cite{Transf_Radio_Maps} to exploit spatial correlations and infer the RSS.

However, transforming the 3D propagation space into a simplified 2D scene inherently curbs the functionality and fidelity of radio propagation tools. First, 2D-based models constrain the prediction to a single receiving plane (\eg at a predefined height). This renders them unsuitable for predicting radio propagation in the 3D space. Second, 3D geometries like buildings or cluttered data are reduced to pixels, hence losing structural details that can produce inaccuracies in the radio propagation modeling.

Alternatively, more robust and realistic representations can be achieved via physics-driven approaches that consider the principles of radio propagation mechanisms. A well-established method is ray-tracing~\cite{Ray_Tracing_Ref}, which simplifies the intricate behavior of wave propagation by assuming that electromagnetic (EM) energy flows along geometrically defined ray paths. These paths are created by the consecutive interaction of a ray/wave emitted from a transmitter with objects found within the propagation environment. Specifically, when a ray impinges onto an interface between two media, it can be partially reflected, refracted, diffracted, or scattered, and there exist laws and approximations to compute the strength and phase of each component \cite{Fields_and_Waves_Cheng}.

In this paper, we present \name, a \textit{graph-driven} framework inspired by the principles of ray tracing that estimates wireless signal-related QoIs in 3D radio environments. The proposed framework directly integrates the radio scene and the ray paths into a unified graph representation, forming the basis of RAN DTs. To do that, we initially build a point cloud representation that captures all objects, transmitters, and receivers in the radio propagation scene. Then, these elements are used to construct a graph where point cloud elements are mapped to nodes in the graph, and edges interconnect all the scene elements by simulating rays (see schematic representation in Figure~\ref{fig:sys_archi}). Particularly, our work yields the following contributions:

\begin{figure*}[!t]
  \centering
\includegraphics[width=0.85\linewidth]{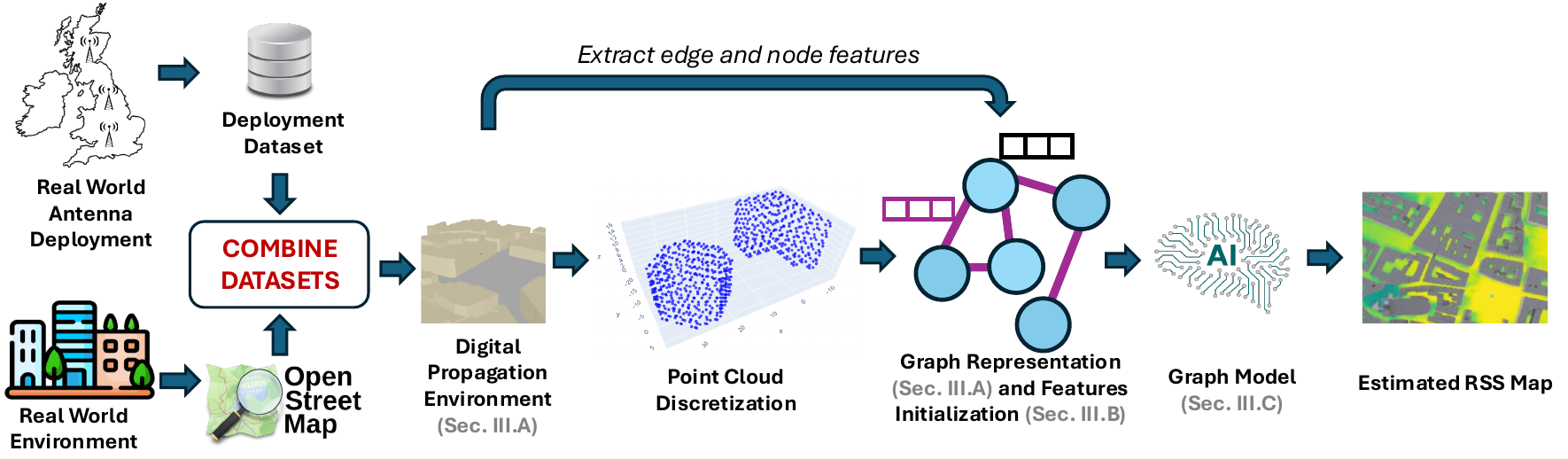}
  \caption{System architecture overview of the proposed method, using various data modalities to construct a DPE, extract an equivalent graph representation, and apply a neural transformation on it to infer RSS.}
  \label{fig:sys_archi}
\end{figure*}

\begin{itemize}

    \item We demonstrate how the digital representation of the propagation environment can be converted into a set of nodes, forming the foundations of our \textit{graph-driven} propagation model. By establishing proper connections between the graph nodes, we instate a representation of the input data that innately resembles the potential paths EM waves can traverse. Unlike the widely adopted simplified 2D visualization \cite{CNN_Propagation, Radio_UNet, EM_DeepRay, GNN_Outdoor, GNN_urban_reconstruction}, our model favors from its physics-inspired architecture.
    
    \item \rrm{We design a set of custom node and edge features that enable our solver to learn to emulate radio propagation through neural message exchange. }
    
    \item We showcase the potential of \name to learn from simulated and real-world measurement data, proving its reliability in a realistic operational setting using data from a production-grade network.     
\end{itemize}

\section{GraphWave Ideation and Problem Formulation}

 \name aspires to establish an equivalent \textit{graph-driven} representation of the propagation environment and the potential paths that EM waves may traverse. To achieve that, it is necessary to convert the continuous 3D physical world into a discretized digital equivalent space, referred to as the digital propagation environment (DPE). 
 Specifically, the DPE includes three main components: ($i$) a set of physical objects, such as buildings, cars and trees, that interact with the EM waves, ($ii$) a transmitting antenna, $T_x$, and ($iii$)  a set of receivers $R$. Each physical object is represented by a collection of triangular meshes in the 3D space, parameterized by its vertex coordinates (see Section~\ref{subsec:graphrepre}). 

In Figure~\ref{fig:sys_archi} we show a general overview of our entire framework, depicting how we build the DPE and extract a graph representation (see Section~\ref{subsec:graphrepre}). Each point in the cloud accounts for a node, $v$,  in a graph, $G$, that resembles the DPE. By properly arranging the connections between the graph nodes, we can create virtual trajectories that reenact the paths through which EM energy flows. Further, each node and edge connection between them is endowed with physics-based and network-related features (see Section \ref{subsec:features}).

\name amalgamates the \textit{graph-driven} representation with a neural transformation to infer a QoI value, $\hat{y}_r$, at each receiving point $r \in R$. This neural transformation is implemented through the exchange of learnable messages between the graph nodes, and corresponds to identifying a mapping function, $f$, that transforms a graph \textit{G} into $\hat{y}_r$ \cite{gilmer2017neural}. Notably, \name exhibits a schematic equivalence with ray-tracing, \ie graph paths correspond to potential ray trajectories, yet its neural nature allows our model to assimilate knowledge and learn from real-world data. Formally, the operation of \name  is defined as:

\begin{equation}
\hat{y}_r = f(G; \theta),
\end{equation}

\noindent where $\theta$ represents the learnable parameters of \name that regulate the message exchange between the graph nodes. Specifically, \name's parameters are computed such that the difference between the predicted and some ground truth values, $y_r$, is minimized. Note that the approach is general in nature and, as we will demonstrate later, \name can be forged either using synthetic data from high-performance propagation simulations or real-world data.

\begin{table*}
\caption{Table with the node and edge features. 
}
\vspace{-2mm}
\label{tab:feature_table}
\centering
\begin{tabular}{c|p{10cm}|c}
\hline
Feature & Description & Node/Edge Feature \\
\hline
    Node type  &  One-hot vector encoding indicating the node type, \ie building, transmitter,  or receiver  & Node \\
    
    $\alpha\_node$  & Angle between the node orientation and the $XY$ plane  & Node \\
    
    $\beta\_node$  & Angle between node orientation and antenna orientation  & Node \\

    Height & Node elevation (in meters) & Node \\
    
    Edge type  & One-hot vector encoding indicating the edge type, \ie transmitter-building, transmitter-receiver and building-receiver  & Edge \\
    
    Distance  & Distance between nodes (in meters)  & Edge \\
    
    SinKR  &  Phase shift real part, np.sin((2*$\pi$/ $\lambda$)*Distance), where $\lambda$ is the wavelength  & Edge \\
    
    CosKR  & Phase shift imaginary part, np.cos((2*$\pi$/$\lambda$ )*Distance), where $\lambda$ is the wavelength   & Edge \\
    
    Free Space Path Loss & Free Space Path Loss value computed using the distance between nodes & Edge \\
    
    $\alpha\_edge$ & Angle between the edge connecting the points and the \textit{XY} plane & Edge \\
    
    $\beta\_edge$ & Angle between the edge and the antenna orientation. For the building-coverage edges, the angle is between the edge and the building node orientation & Edge \\
    
    Radiation Pattern & Only for transmitter-receiver and transmitter-building edges. It stores the gain value where the edge intersects the radiation pattern of the transmitter & Edge \\
\hline
\end{tabular}
\vspace{-0.2cm}
\end{table*}

\section{Model Design}
\label{sec:model_design}

In the following subsections, we present the design choices and implementation details of the components that constitute \name. Specifically, we discuss how we generate the DPE and the equivalent graph representation, the selection of the node and edge input features, and the operation of the graph neural network (GNN) used to predict $\hat{y}_r$.  

\subsection{Digital Propagation Environment $\&$ Graph Representation}
\label{subsec:graphrepre}

Each DPE comprises essential environmental and network-related details, including the object layout and its properties, wireless device locations, and the respective transmitter antenna configuration. The 3D objects in the DPE are decomposed into a set of distinct triangular meshes, with each triangle capturing a portion of the object's surface. The union of all triangular meshes precisely recreates the shape of each structure, ensuring an accurate geometric representation of the propagation environment. \rrm{This is a common practice in computer vision since any 3D surface can be represented by a mesh of triangles, which are simpler and they have consistent geometric properties.} We note that in this work, the only type of objects considered are buildings, however, the outlined methodology is general and can also be applied to model other entities, such as cars.

Subsequently, the triangular meshes are utilized to derive a point cloud representation of the DPE. This is achieved by iterating over all triangles within the DPE, and for each one of them we create a discrete 2D plane with uniform spacing between its points, eventually retaining in the cloud only the points that overlap with the triangle surface. In addition to digitizing physical objects, the wireless equipment, \ie the transmitting antenna $T_x$ and the set of receivers $R$, are also digitized and represented as distinct points of the cloud. These are placed accordingly to their relative position to the buildings in the real-world.

The constructed cloud point sets the basis for constructing the graph representation of the DPE. To do this, the cloud points are considered distinct nodes, and they are interconnected to form a graph $G_r = (V_r, E_r)$ for each receiver $r \in R$, where $V_r$ is the set of nodes and $E_r$ is the set of edges. The edges are created by connecting the transmitter node to all building nodes, and then linking all the building nodes with the receiver node. In addition, we add a distinct edge from the transmitter to the receiver to consider the case of line-of-sight propagation. 
\rrm{Note that this connectivity policy does not explicitly reconstruct the ray-tracing paths, but instead provides a graph-based approximation of all potential propagation routes from the transmitter to the receiver. However, in its current format, this representation is limited to first-order reflections, and we leave the inclusion of additional edges that will capture higher-order reflections as future work.}

\subsection{Features Initialization}
\label{subsec:features}

To enable \name to learn the EM propagation dynamics, each node and edge must be assigned with some input features that will be processed neurally to derive $\hat{y}_r$. Specifically, each node $v \in V_r$ is associated with a feature vector $\mathbf{x}_v \in \mathbb{R}^{d_V}$, forming the node feature matrix $\mathbf{X}_V \in \mathbb{R}^{|V_r|\times d_V}$. There are three types of nodes in the graph: transmitter nodes, building nodes, and receiver nodes. Each node feature vector is initialized with a one-hot encoding vector to indicate its type. Additionally, each node is assigned an orientation to represent its spatial direction. In particular, for a transmitter node the orientation is determined by its azimuth and tilt angles, whilst all the receivers $R$ are oriented towards the positive \textit{Z}-axis. Lastly,  the orientation of each building node is defined perpendicular to the 3D triangle that is part of. These orientations are encoded using two angles: $\alpha_v$ and $\beta_v$. The first represents  the angle between the node orientation and the \textit{XY} ground plane, and the latter depicts the angle between the node orientation and antenna orientation. Note that this design choice allows our \textit{graph-driven} neural propagation solver to be rotation equivariant ~\cite{satorras2021n}.

Similarly, each edge $e \in E_r$ has an associated feature vector $\mathbf{x_e} \in \mathbb{R}^{d_E}$, which defines the edge feature matrix $ \mathbf{X_E} \in \mathbb{R}^{|E_r|\times d_E}$. The edge features capture relationships between nodes, including distance, incidence angle, and visibility. There are three categories of edges: transmitter–building, building–receiver, and transmitter–receiver. Each edge is initialized with a one-hot encoding vector to represent its category. In addition, a set of supplementary features is concatenated to the edge feature vector. Table~\ref{tab:feature_table} shows a summary of all the features used to initialize the node and edges feature vectors. Similarly to the nodes angle features, we also assign an edge angle orientation $\alpha_e$ and $\beta_e$.

\subsection{Graph Neural Network Model}
 
 The graph representation, along with the carefully designed features, enable portraying faithfully the geometric characteristics of the propagation environment. Further, the selected features provide meta-information that helps capture properties related to the antenna orientation with respect to the receivers and the buildings, as well as how potential path/edges affect the propagation of EM waves, \eg phase shift and path attenuation. Bringing these pieces of information together through a unified neural operator, \name learns and quantifies the impact of various propagation mechanisms on EM wave strength, such as reflection, diffraction, and scattering.

 The neural operator is implemented via a GNN that adheres to the neural message passing process outlined in~\cite{gilmer2017neural}, estimating $\hat{y}_r$ at the receiving nodes through $L$ consecutive message-passing operations and a final readout stage. During the message-passing phase, each node feature vector, $\mathbf{x_v}$, is transformed into a hidden state, $\mathbf{h_v}$, based on the information gathered from neighboring nodes and the respective connecting edges. For the $l+1$-th message passing operation, this transformation is achieved  via two learnable non-linear functions, $f_m$ and $f_u$, parameterized via neural networks and employed to evaluate the transmitted messages, $\mathbf{m}_v$, and update the node hidden states, respectively: 

\vspace{-3mm}
    \begin{equation}
      \mathbf{m}_{v_i}^{(l+1)}  = \sum_{j \in  N(v_i)  } f_m(  \mathbf{h}_{v_i}^{(l)} , \mathbf{h}_{v_j}^{(l)},  \mathbf{x_e}^{{i,j}}  )      ,  
    \label{eq:MPNN_message}
\end{equation}

\vspace{-2mm}
\begin{equation}
      \mathbf{h}_{v_i}^{(l+1)}  =   f_u(\mathbf{h}_{v_i}^{(l)},  \mathbf{m}_{v_i}^{(l+1)}),   
    \label{eq:Update}
\end{equation}

\noindent \textcolor{black}{where  $\mathbf{x_e}^{{i,j}}$ is the edge feature vector of the edge connecting the $i$-th and the $j$-th nodes, whilst $\mathbf{h}_{v_i}$ and $\mathbf{h}_{v_j}$ are the hidden state feature vectors for the $l$-th  message-passing layer (note that for the zeroth layer these corresponds to $\mathbf{x}_{v} )$}. Then, in the readout phase, $\hat{y}_r$ is estimated  through a neural readout function, $f_r$, as: 
\vspace{-4mm}

\begin{equation}
      \mathbf{\hat{y}_r}  =   f_r(   \mathbf{h}_{v_i}^{(L)} | v \in V_r ).   
    \label{eq:Readout}
\end{equation}

\noindent and  the loss function to be minimized is  the mean absolute error (MAE) between the prediction and the ground truth values:

\vspace{-3mm}

\begin{equation}
\mathcal{L}(\theta) = \frac{1}{|R|} \sum_{r \in R} \left|\hat{y}_r - y_r \right|
 \label{eq:loss}
\end{equation}

\section{Evaluation}

In the following subsections, we discuss the details of the dataset used to forge \name. We provide a proof-of-concept for our \textit{graph-driven} neural propagation model, showcasing how it can replicate the results of a ray-tracer. We then exemplify its potential to learn from real-world data and outperform conventional propagation simulators. To finalize, we discuss the open challenges for building AI-based radio propagation tools. \rrm{All the experiments where executed on off-the-shelf hardware with an Intel(R) Xeon(R) Gold 6348 CPU with 2.60GHz and an NVIDIA RTX A5000 GPU.}

\subsection{Datasets}
\label{subsec:dataset}

\name is built using three different data sources, conveying information related to the network configuration, the propagation environment layout, and the target QoI. The first data source comes from a major MNO in Europe that comprises real-world antenna placements and their configuration, \ie operating frequency, azimuth and tile angles, beam width. This information is then used in conjunction with data from OpenStreetMap (OSM)~\cite{OpenStreetMap} to extract 3D building geometric information and construct the DPEs. To do this, we create a bounding box that surrounds the antenna (\ie $200 \times 200$ meters with the antenna placed in the middle), and download all the geometric information for all the buildings within this area. Subsequently, we use Blender 4.0.0~\cite{BlenderWebsite} and Python 3.10.12 to iterate over the triangular mesh and convert this geometric information into a point cloud, as per Section~\ref{subsec:graphrepre}. The triangular mesh is implicitly defined in the OSM dataset, hence no additional step is required to obtain the triangles.

The third dataset includes the ground truth values that will be used to train \name, which can be obtained either through software simulations or real-world measurements from the operating network. In Section~\ref{subsec:proofconcept} we used simulated data generated via Sionna RT 0.19.0 \cite{10465179} in order to verify that indeed \name can produce results commensurate to that of existing propagation solvers. Yet, the advantage of \name is that due to its neural nature, it can accrue knowledge from real data. To highlight this strength, in section~\ref{ref:exp_empirical} we used crowdsourced data from a major MNO in Europe. Using the latitude and longitude of each measurement allowed us to place them within our DPE.

\subsection{Proof-of-Concept with Ray-Tracing Data}
\label{subsec:proofconcept}

\begin{figure}[!t]
\centering
    \subfigure[]
    {\includegraphics[width=0.470\columnwidth]{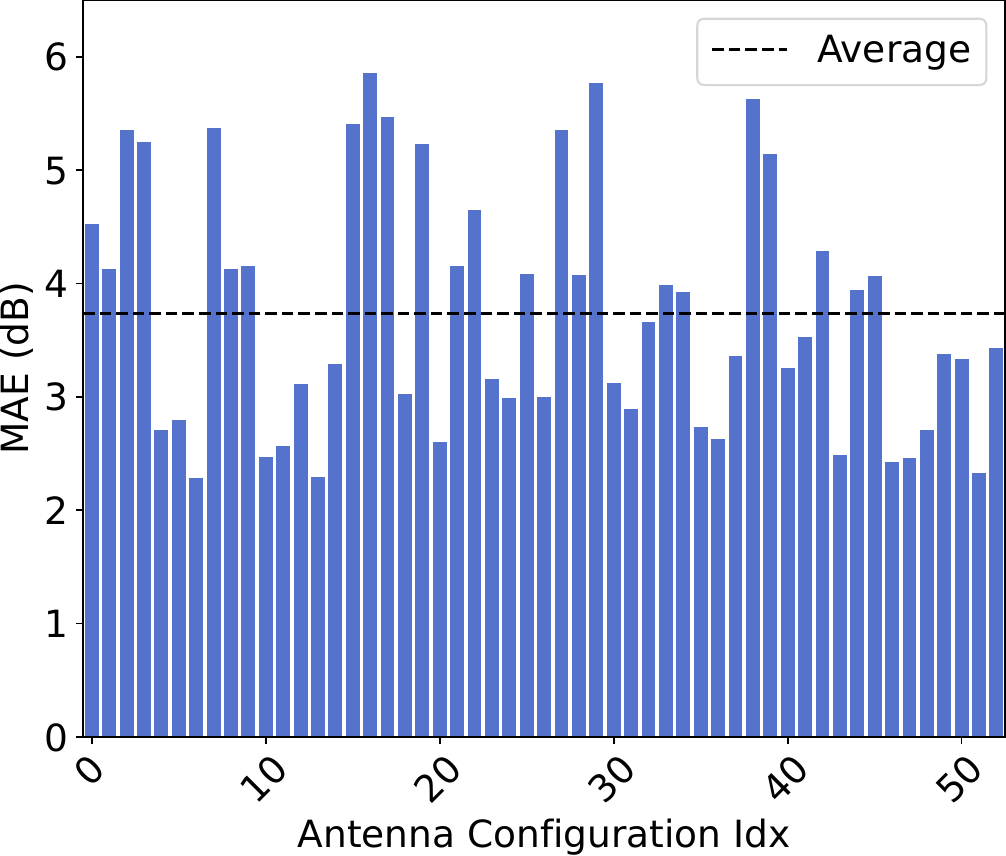}
    \label{fig:mae_barplot}}
    \subfigure[]
    {\includegraphics[width=0.470\columnwidth]{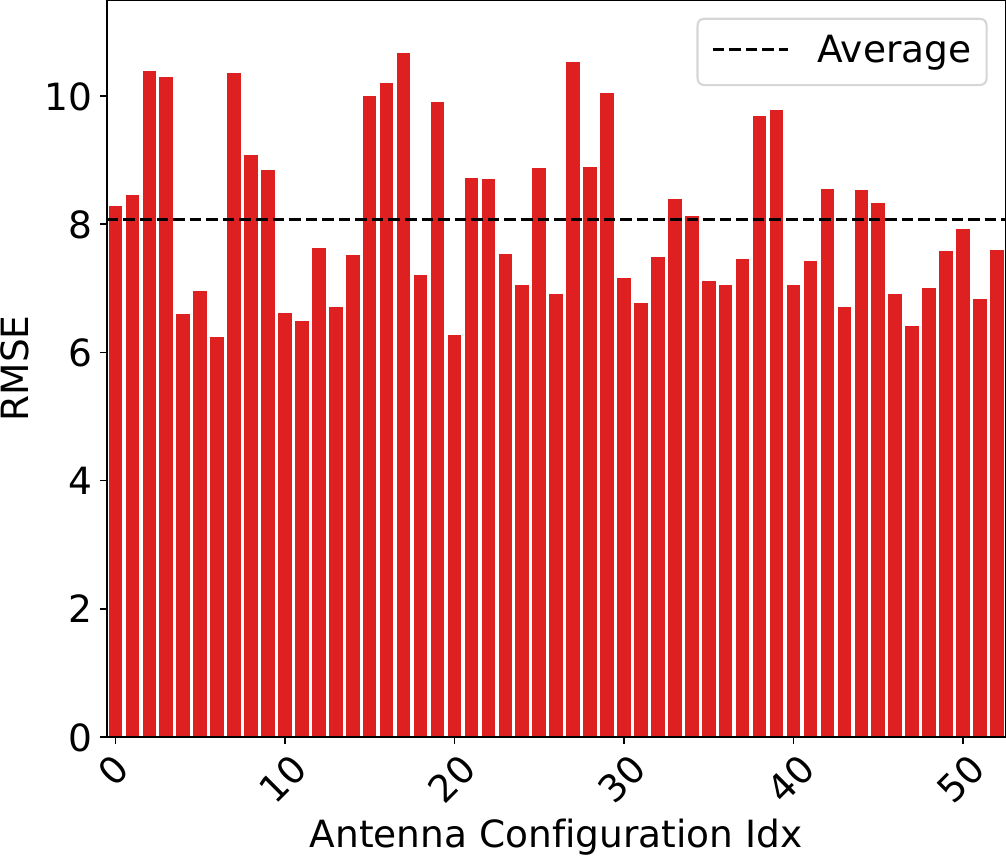}
    \label{fig:rmse_barplot}}
    \caption{(a) MAE and (b) RMSE sample distribution between \name and the ground truth values obtained with Sionna RT. The $x$-axis indicates the antenna configuration index from the 53 different antenna configurations used for testing.}
    \label{fig:barplot}
\vspace{-0.3cm}
\end{figure}

The objective of this set of experiments was to demonstrate the capabilities of \name to predict the RSS. \rrm{Specifically, we aim at observing the gap between the proposed model and the ground truth obtained with Sionna RT.} To this end, we took a real-world antenna configuration, and we built a DPE of 200x200 meters using OSM. Then, we created a synthetic dataset of 71 different antenna configurations, where we changed the original azimuth and tilt angles. Specifically, the tilt angle ranges from -10 to +15 in steps of five and these values were added to the original tilt value from the antenna configuration. Similarly, for the azimuth we explored a range from -25 to +30 that was added to the original value. For each antenna configuration, we used the Sionna RT~\cite{10465179} version 0.19.0 to launch the simulations and obtain the ground truth RSS values. Sionna was configured with a cell size of (2, 2) meters, with a maximum reflection depth of 8, with $1\cdot10^{6}$ rays traced and with diffraction enabled. 

From a total of 71 pairs of different tilt and azimuth values, we selected 18 of them for training and 53 for testing. For each pair, we built a DPE, and we placed a set of receivers so that we could use Sionna RT to simulate the RSS values. During training, we randomly selected 70\% of the receiver positions for training and the remaining 30\% were used for validation,  which is a common dataset split ratio in machine learning. We set a size of 64 for the feature vectors of the nodes and edges, and we used 256 neurons in the two-layer of the readout with a dropout of 0.1 and a batch size of 32. The learning rate remained constant at 0.001 during the entire training process of 3000 epochs. The model with the lowest validation loss is selected for further evaluation in the test split.

\begin{figure}[!t]
\centering
    \subfigure[]
    {\includegraphics[width=0.470\columnwidth]{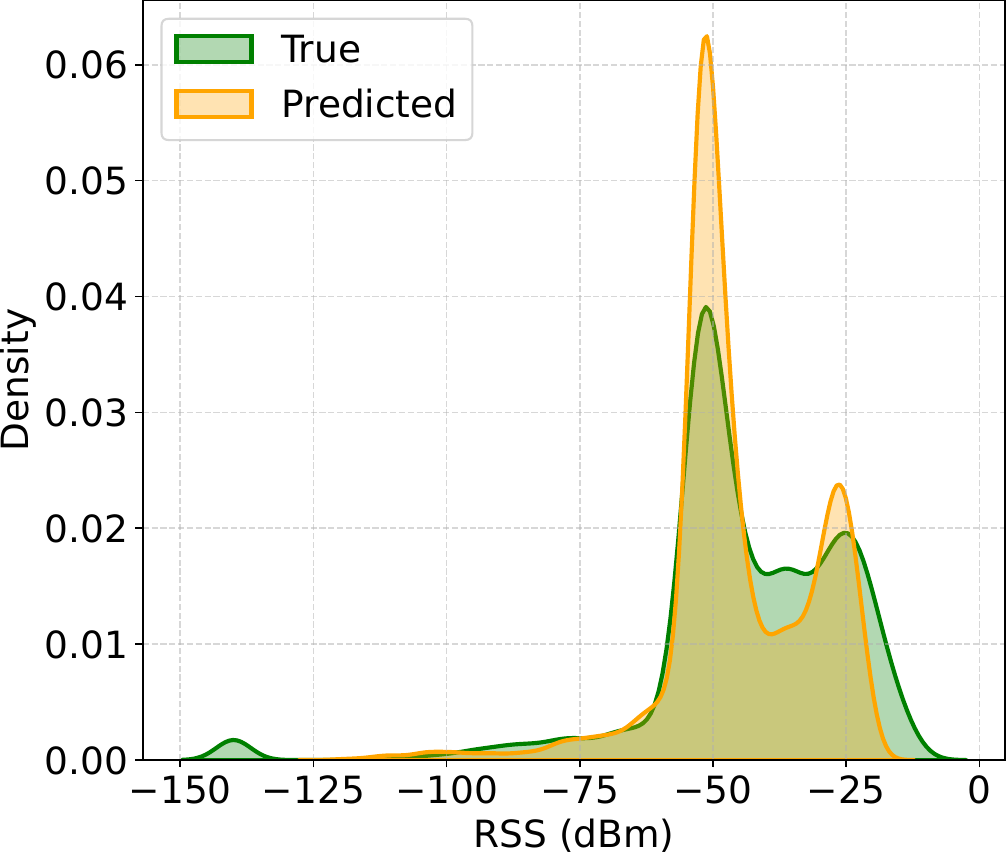}
    \label{fig:max_mae}}
    \subfigure[]
    {\includegraphics[width=0.470\columnwidth]{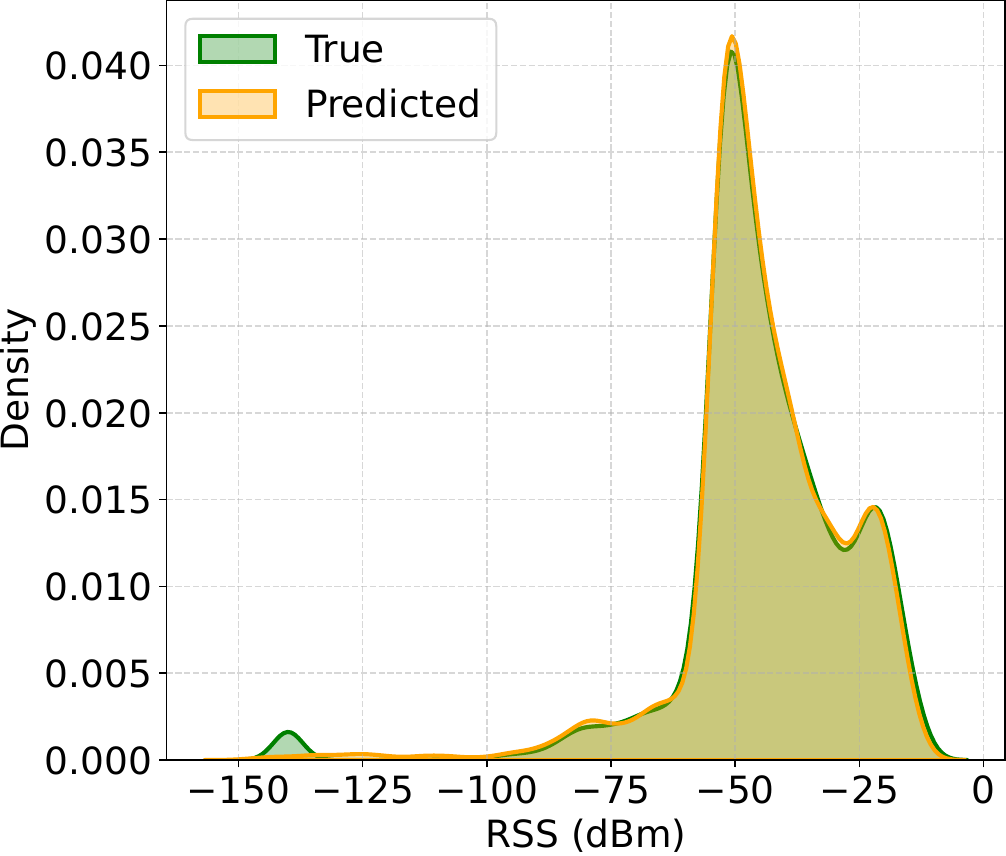}
    \label{fig:min_mae}}
    \caption{Continuous probability density curves for the predictions in the antenna configuration index with (a) maximum MAE and (b) minimum MAE in the test set.}
    \label{fig:distribution_plot}
\vspace{-0.2cm}
\end{figure}

Figure~\ref{fig:barplot} shows the mean absolute error (MAE) and root mean squared error (RMSE) between \name's predictions and the ground truth RSS values for the test set. \rrm{This comparison allows us to measure the gap between the GNN-based model and the ground truth values simulated with Sionna RT.} \name achieves an MAE of 3.73 dB with a range of values from 2.28 to 5.86 dB, demonstrating robust performance throughout different antenna configurations than those seen during training. In terms of execution cost to obtain the RSS maps, we observe an average time of 1.03 seconds for \name and 1.86 seconds for Sionna RT. This gap, even though it's small, indicates a potential for GNN-based methods to make fast, accurate predictions. 

\rrm{From the experiments described above, we selected the antenna configuration index with highest and lowest MAE in the test set to visually compare the distributions of the predicted values by \name with respect to the ground truth obtained with Sionna RT 0.19.0. Figures~\ref{fig:max_mae} and ~\ref{fig:min_mae} show the probability density curves of \name's predictions together with the true values for the DPE with maximum and minimum MAE respectively. For the case with the highest MAE, the results indicate that the distribution of predicted values does not fully match the ground truth values. Nevertheless, the peaks in both distributions coincide, indicating the potential of \name to learn the key radio propagation aspects for the DPE. Alternatively, the case with minimum MAE from Figure~\ref{fig:min_mae} shows the distributions are almost identical. The difference lies in the non-coverage values with an RSS of $\approx$ -140 dBm, indicating that the deviations between \name and ray-tracing are not critical. }

\begin{figure}[!t]

\centering
    {\includegraphics[width=0.70\columnwidth]{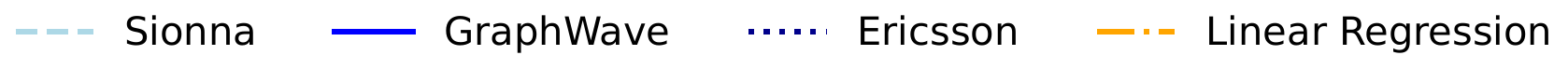}}
    \subfigure[]
    {\includegraphics[width=0.85\columnwidth]
    {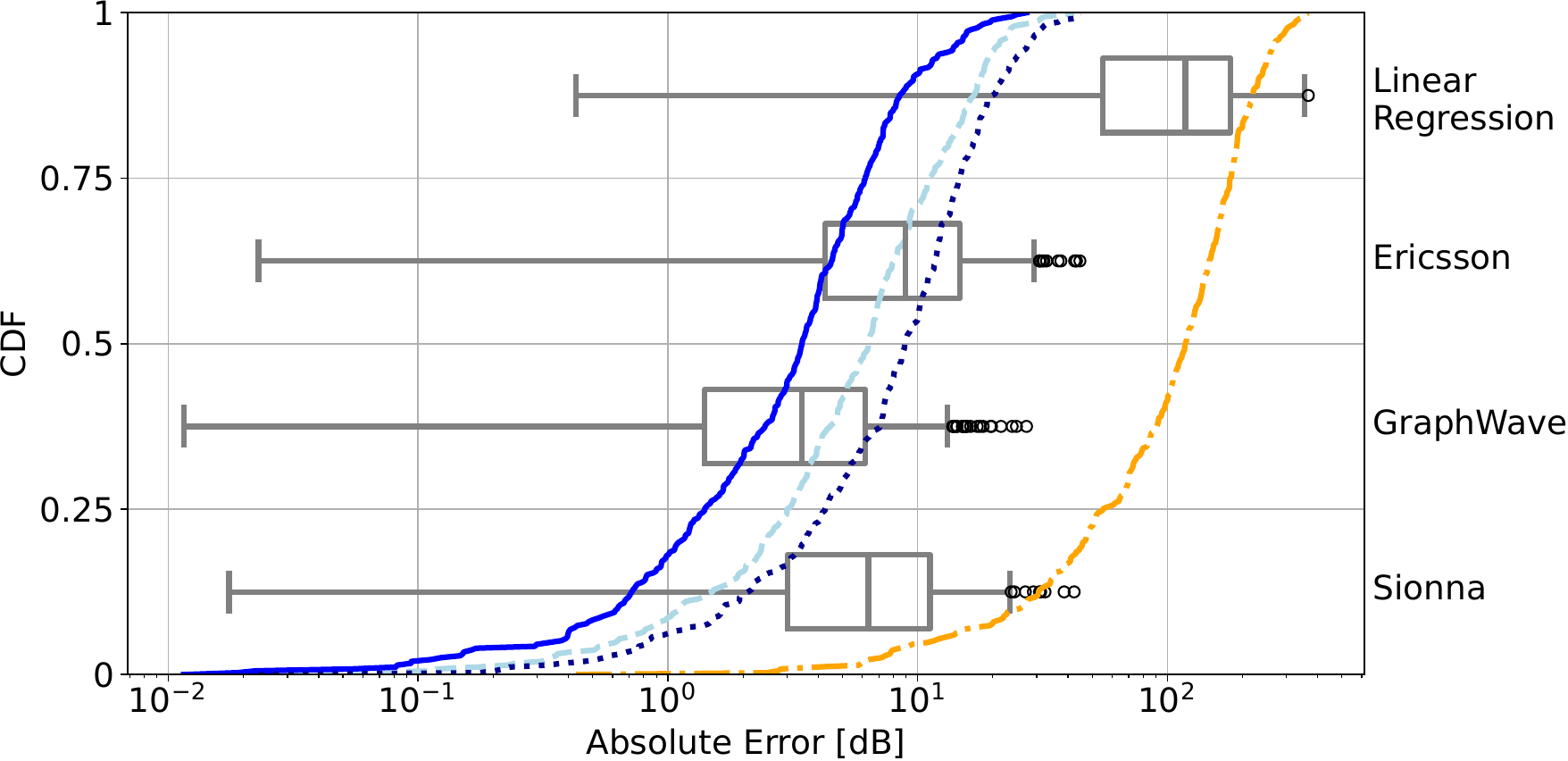}
    \label{fig:empirical_results_oxford}}
    \subfigure[]
    {\includegraphics[width=0.85\columnwidth]{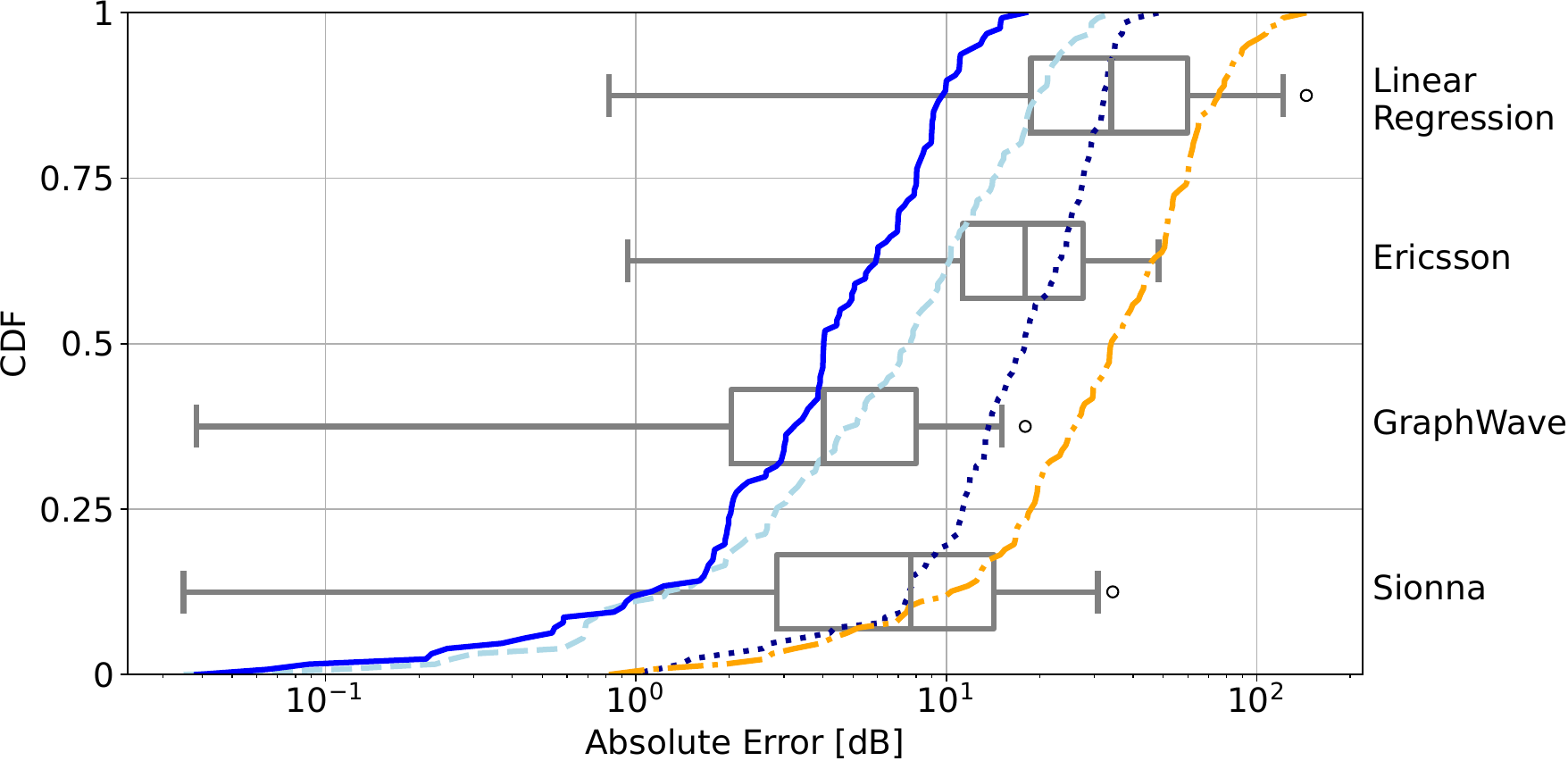}
    \label{fig:empirical_results_cambridge}}
    \caption{Results with real-world measurements on (a) city A and (b) city B.}
\vspace{-0.2cm}
\end{figure}

\subsection{High-fidelity RSS Inference with Real-World data}
\label{ref:exp_empirical}

In this section, we exemplify the potential of \name to generate high-fidelity empirical coverage maps from real-world data. Traditional simulations often fail to accurately represent real-world environments due to the difficulty of gathering critical meta-information, such as the construction material properties. Not having this kind of information or making assumptions about it, \eg assuming all buildings are made of concrete, along with the innate limitations of the propagation solver, results in discrepancies between the obtained simulated values and the empirical measurements. In this set of experiments, the objective was to train \name with real-world measurements and use the trained model to obtain realistic high-fidelity estimations. Since measurements capture radio propagation in the real-world, with \name we are reducing the natural gap between simulations and real-world radio propagation.

In these experiments, we leveraged a real-world crowdsourced measurements dataset from a MNO in Europe (see Section~\ref{subsec:dataset}). Initially, we selected the RSS measurements from two different cities, referred to as city A and B, and we merged them with the antenna deployment dataset, allowing us to associate the measurements with the antennas they are connected to. For each of these antennas, we had their configuration and radiation pattern, which allowed us to faithfully represent the antenna configuration in the DPE. We selected one antenna from city A and one from city B, creating two different DPEs with a set of empirical measurements placed on the respective scenes as they were in the real-world. The empirical measurements were randomly split into 70\% of them for training, 20\% for validation and 10\% for testing. We trained \name, and we evaluated its performance on the testing set for each  DPE independently. As baselines, we used a Linear Regression, Sionna \cite{10465179} and the empirical Ericsson channel model, which is a  modification of Okumura-Hata model with tunable parameters set according to \cite{Ericsso}.

Figure~\ref{fig:empirical_results_oxford} shows the evaluation results for city A. The x-axis corresponds to the absolute error between the real-world empirical measurements and the estimated RSS values for each model. On the y-axis on the left-hand side, we can read the cumulative density function (CDF) of the same errors. The experimental results indicate that $\geq$90\% of the samples predicted by \name have an absolute error between 0 and 10 dB. In addition, \name has a mean error of -0.03, whereas Sionna has a mean of 3.51 and Ericsson and Linear Regression models have a mean error of -6.7 and 113.46 respectively. These values indicate that Sionna and Linear Regression tend to underestimate the RSS values and the Ericsson model overestimates them on average. In Figure~\ref{fig:empirical_results_cambridge} we show the results for city B having similar trends as in city A. The means are 0.46, -19.16, 0.39 and 3.28 for \name, Ericsson, Linear Regression and Sionna respectively. Similarly, in City A, Sionna and Linear Regression underestimate the RSS, the Ericsson  model overestimates the predictions and \name offers more equilibrated predictions.

\subsection{Discussion}

The experimental evaluation showed the potential of \name to learn radio propagation in 3D environments and estimate QoIs for a given area of interest. Even though the results are promising, there are several open challenges that need to be addressed to mold a generic AI-powered radio propagation tool. We envision a tool with high performance and low computational complexity regardless of the 3D environmental characteristics or the transmitter configuration. To do this, we need to solve the lack of real-world datasets with accurate geometric meta-information of the buildings and their materials. In our work, we used OSM as a data source for this information, but we noticed that the building heights are approximated and the building materials information is, in most cases, incomplete. This naturally results in discrepancies between the simulated scenarios and the real-world environment.

The implementation of \name relies on representing the DPE as a point cloud, a design choice that was inspired by the EM wave propagation in a 3D environment. We acknowledge that this representation limits the scalability of our method to substantially larger DPEs due to the limited GPU memory needed to train the GNN. \rrm{As future work, we are planning to analyze techniques like subgraph chunking and other mechanisms to handle larger DPEs while reducing graph size and memory consumption. In addition, we are planning to extend \name by incorporating additional propagation mechanisms such as scattering or second-order reflections that would lead to more accurate modeling.}

\section{Conclusions}

In this paper, we presented a \textit{graph-driven} neural radio propagation modeling approach inspired by ray-tracing. We demonstrated how to extract a point cloud representation from a digital replica of a real-world propagation environment, and then convert it to an equivalent graph with features in the nodes and edges. Consequently, we assigned proper physics-based and system-related features to the graph nodes and edges that were effectively transformed to an end-user RSS value via neural message passing. Our experimental results indicated that \name can learn from both synthetic and real-world data. The proposed framework showed better accuracy compared to ray-tracing for real-world datasets, paving the way for the establishment of physics-inspired tools that can be exploited by MNOs for efficient RAN optimization and predictive performance analysis.

\section*{Acknowledgment}

This work was supported by the SNS JU under the European Union’s Horizon Europe research and innovation program under Grant Agreements No. 101139161 (INSTINCT) and No. 101192369 (6G-MIRAI). Views and opinions expressed are however those of the authors only and do not necessarily reflect those of the European Union or SNS JU. Neither the European Union nor the granting authority can be held responsible for them. Stefanos Bakirtzis was supported by the Foundation for Education and European Culture.

\printbibliography

\end{document}